\documentstyle[12pt,aaspp4]{article}

\def\h2{H$_2$}
\def\halpha{H$\alpha$}
\def\microns{$\mu$m}
\def\brgamma{Br$\gamma$}
\def\msun{M$_\odot$}
\def\lsun{L$_\odot$}
\def\etal{et al.\ }

\begin{document}
\onecolumn
\title{The Global Schmidt Law in Star Forming Galaxies
\footnote{Visiting Astronomer, Kitt Peak National Observatory, National
Optical Astronomical Observatories, which are operated by the Association
of Universities for Research in Astronomy, Inc., under contract with
the National Science Foundation.}
}

\author{Robert C. Kennicutt, Jr.}

\affil{Steward Observatory, University of Arizona, Tucson, AZ 85721}

\begin{abstract}

Measurements of \halpha, HI, and CO distributions in 61 normal spiral 
galaxies are combined with published far-infrared and CO observations
of 36 infrared-selected starburst galaxies, in order to study the form of the
global star formation law, over the full range of gas densities and
star formation rates (SFRs) observed in galaxies.  The disk-averaged SFRs 
and gas densities for the combined sample are well represented by a Schmidt
law with index $N = 1.4 \pm 0.15$.  The Schmidt law provides
a surprisingly tight parametrization of the global star formation law,
extending over several orders of magnitude in SFR and gas density.  
An alternative formulation of the
star formation law, in which the SFR is presumed to scale 
with the ratio of the gas density to the average orbital timescale, also
fits the data very well.  Both descriptions provide
potentially useful ``recipes" for modelling the SFR in numerical simulations of
galaxy formation and evolution.

\end{abstract}


\section{INTRODUCTION}

A key ingredient in the understanding and modelling of galaxy evolution
is the relationship between the large-scale star formation rate (SFR)  
and the physical conditions in the interstellar medium (ISM).  Most current 
galaxy formation and evolution models treat star formation using simple  
ad hoc parametrizations, and our 
limited understanding of the actual form and nature of the 
SFR-ISM interaction remains as one of the 
major limitations in these models (e.g., Navarro \& Steinmetz 1997).
Measurements of the star formation law in nearby 
galaxies can address this problem in two important respects, by providing 
empirical ``recipes" that can be incorporated into analytical models and 
numerical simulations, and by providing clues to the physical mechanisms that 
underlie the observed correlations. 

The most widely applied star formation law remains the simple gas density 
power law introduced by Schmidt (1959), which for external galaxies is 
usually expressed in terms of the observable surface densities of gas and 
star formation: 
\begin{equation} 
\Sigma_{SFR} = A~\Sigma{_{gas}^N} 
\end{equation}

\noindent
The validity of the Schmidt law has been tested in dozens of 
empirical studies, with most measured values of $N$ falling in the range 
1 $-$ 2, depending on the tracers used and the linear scales considered 
(Kennicutt 1997).  On large scales the star 
formation law shows a more complex character, with a Schmidt law at high 
gas densities, and a sharp decline in the SFR below a critical threshold 
density (Kennicutt 1989, hereafter K89).  These thresholds appear to be 
associated with large-scale gravitational stability thresholds for massive 
cloud formation (e.g., Quirk 1972; Fall \& Efstathiou 1980; K89).  
At high gas densities, well above the 
stability threshold, the form of the Schmidt law appears to be remarkably 
consistent from galaxy to galaxy, both in terms of its slope ($N \sim 1.3 
- 1.5$) and the absolute SFR efficiency (the coefficient $A$ in eq. [1]).  
Studies of this kind 
offer the beginnings of a quantitative, physical prescription for the SFR 
that can be incorporated into galaxy formation and evolution models.

This is the first of two papers which reinvestigate the form and physical 
nature of the star formation law, over a much larger range of galaxy types 
and gas densities than was possible previously.  Paper II (Martin \& 
Kennicutt 1998) uses new \halpha\ CCD imaging of an HI and CO 
selected sample of spiral galaxies to quantify the behavior of the 
star formation law within individual galaxies, and to test several 
models for the star formation law.  This paper is 
concerned with the behavior of the star formation law on global scales, 
averaged over the entire star forming disk.  Such global laws, which treat 
galaxies in a single-zone approximation, provide less physical insight 
into the star formation process itself, but they provide very useful 
parametrizations (recipes) for galaxy evolution modelling.  

Earlier work has shown that the global, disk-averaged star formation 
law is reasonably well represented by a Schmidt law (K89; Buat, Deharveng, 
\& Donas 1989; Buat 1992; Boselli 1994; Deharveng \etal 1994).  However
these analyses have been hampered by small samples and by the small  
range of gas densities represented in those 
samples.  In this paper we use newly 
available HI, CO, and \halpha\ data to more than double the sample over
previous studies, and fully cover the range of mean gas 
densities found in disks.  We combine these data with published CO, 
\brgamma, and far-infrared (FIR) measurements of luminous 
starburst galaxies, to investigate the nature of the Schmidt law in  
higher density environments, thereby 
extending the total density range probed to nearly five orders of 
magnitude.  Our main goal is to  
test whether the millionfold range in observed SFRs,  
extending from quiescent gas-poor disks to nuclear starbursts, 
can be understood within a common empirical and physical framework. 

% The remainder of this paper is organized as follows.  The data compiled
% for the samples of normal galaxies and starbursts are presented and 
%discussed separately in \S 2.  In \S 3 we analyze the form of th edisk-averaged 
% Schmidt law for normal star forming disks and starburst galaxies separately.
% In \S 4 we discuss the composite 
% star formation law, and interpret it in terms of a traditional Schmidt law 
% as well as in terms of a simple dynamical timescale model (e.g., Silk 
% 1997).  We find that either description provides a very useful 
% parametrization for the global SFR, that appears to hold over the entire 
% range of star formation scales that is observed in local galaxies.

\section{DATA}

To investigate the global star formation law in normal disks, we searched 
the literature for normal galaxies with well-sampled HI and CO 
measurements, and for which \halpha\ imaging or photometry are available.  Our 
analysis of this sample closely follows that described in K89.  To investigate 
the star formation law at higher densities, we compiled published CO maps, 
FIR photometry, and \brgamma\ emission-line measurements for a sample of 
infrared-selected starburst galaxies.  Each data set is discussed 
separately below.

\subsection{Normal Disks}

Previous studies of the disk-averaged star formation law have shown that 
the global SFR correlates most strongly with the total (atomic $+$ 
molecular) gas density (e.g., Kenney \& Young 1988; K89; Buat 1992; 
Boselli 1994).  Consequently our primary data set consists of normal 
spirals for which spatially-resolved HI, CO, and \halpha\ data are 
available.  A master list of candidate galaxies was compiled from the
FCRAO CO survey (Young \etal 1989; 1995), supplemented by the CO survey of Sage 
(1993).  Within these samples, we identified 61 galaxies which also have 
published HI maps, \halpha\ photometry, and inclinations less than  
75\arcdeg\ (to avoid severe extinction problems in edge-on systems).  
Total HI masses based on single-dish measurements are available for 
another 150 galaxies, but those data are unsuitable for the current 
application, because much of the HI is located well outside of the star 
forming disks, and it is essential to correlate the SFR and gas densities 
over the same physical region.  However we do use some of these additional
galaxies in \S3.1 to examine the form of the SFR {\it vs} HI Schmidt
law.

Table 1 lists the 61 galaxies in the 
sample, the relevant surface densities, and references, as 
described below.  When considering the sample properties as a whole
the main selection criterion was availability of CO and HI maps,
so the galaxies should comprise a virtually unbiased set in terms
of star formation properties.  Approximately 40\%\ of the galaxies
are members of the Virgo cluster, selected from the CO survey
of Kenney \& Young (1988) and the HI survey of Warmels (1988), and
this sample contains most of the luminous spirals in the cluster core.
The field galaxy subsample is more heterogeneous, and is significantly
biased toward galaxies of Hubble type Sb and later, but it is unlikely
that this selection biases the form of the star formation law.

HI surface densities were taken mainly from the compilations of Warmels (1988), 
Broeils \& van Woerden (1994), and Rhee \& van Albada (1995), supplemented by   
individual measurements of a few galaxies (Table 1).  The mean HI surface
densities, averaged within the optical radius of the disk, were
derived from the surface density profiles 
given in those papers or the references therein.  The disk radii 
are the corrected isophotal radii as given in the RC2 catalog 
(de Vaucouleurs, de Vaucouleurs, \& Corwin 1976).  The mean densities
used here differ from those that are often tabulated in the original
papers, the latter usually being
averaged within the inner {\it half} of the optical disk.  

Total molecular hydrogen masses were taken from the Young \etal 
(1989; 1995) and Sage (1993) surveys, and converted when needed to
a common CO/H$_2$ 
conversion factor:  $N(H_2) = 2.8 \times 10^{20}~I_{{\rm CO}}$~cm$^{-2}$~ 
(K~km~s$^{-1}$). 
%and multiplied by 1.30 to include helium and heavy elements. 
The mean \h2\ surface densities were then determined, by averaging within the 
radii listed in Table 1.  These average densities are meaningful only if 
the CO emission is confined to the optical disk, and the measurements extend  
to a substantial fraction of optical radius.  Galaxies  
which were sampled to less than half of the optical radius were not  
included in our sample.

Integrated SFRs were derived from measurements of the \halpha\ 
emission-line flux, following the method described in Kennicutt (1983).  
Most of the \halpha\ fluxes were taken from the surveys of Kennicutt \& 
Kent (1983), Romanishin 
(1990), and Young \etal (1996).  Those data were supplemented with new 
calibrated \halpha\ CCD images obtained with a focal reducer camera on the 
Steward Observatory 2.3~m Bok telescope, and with the 0.9~m and Burrell Schmidt 
telescopes at Kitt Peak National Observatory.  Details of these 
observations are given in Paper II.  The \halpha\ 
fluxes were corrected as needed for foreground extinction and [NII] 
emission, following the prescriptions in Kennicutt (1983).  The  
original \halpha\ fluxes of Kennicutt \& Kent (1983) have been corrected
upwards by a factor of 1.16 to place them on a consistent zeropoint with
more recent measurements (Romanishin 1990; Kennicutt 1992).  

The \halpha\ luminosities were then converted to total SFRs,  
using the updated calibration 
of Kennicutt, Tamblyn, \& Congdon (1994): 
\begin{equation} 
SFR~({\rm M_\odot~yr^{-1}}) = {{L(H\alpha)} \over {1.26 \times 10^{41} 
~{\rm ergs~s^{-1}}}} 
\end{equation}

\noindent 
The \halpha\ luminosities used in equation (2) were corrected 
for internal extinction by 1.1 mag (factor 2.8), based on a 
comparison of free-free radio fluxes and \halpha\ fluxes of 
galaxies by Kennicutt (1983).   
The actual extinction varies within the sample, of course, 
which introduces significant scatter in the observed star 
formation law, as discussed later.  While it would be much
better to apply individual extinction corrections to each galaxy,
determining the reddening or extinction from integrated spectra
is problematic (Kennicutt 1992), and would introduce uncertainties
that are larger than the single average correction.  It may be
possible in the future to derive improved estimates of the extinction
and SFR using measurements of near-infrared Brackett or Paschen
recombination lines, but such data are not currently available.

The IMF used in this conversion is a Salpeter function  
($dN(m)/dm = -2.35$) over $m = 0.1 - 100$ \msun.  The Salpeter IMF was
adopted in order to be consistent with the infrared-derived SFRs in the
next section.  Adopting the extended Miller-Scalo function
used in Kennicutt (1983) would produce nearly identical SFRs (only 8\%
lower). Galaxy distances from Young \etal (1989) were used in this intermediate 
calculation, but the distances are irrelevant for most of this paper, 
because the Schmidt law is analyzed in terms of 
distance-independent surface densities.  

Finally, the mean SFR surface density (units \msun~yr$^{-1}$~kpc$^{-2}$) was 
derived for each galaxy, by dividing the total SFR from equation (2) by
the deprojected area within the corrected RC2 radius.
Through the remainder of this paper, we shall refer to this SFR per unit
area as the ``SFR density".   In most galaxies
the RC2 radius coincides approximately with the edge 
of the main \halpha-emitting disk (K89), so the SFR density as measured here
corresponds roughly to the mean SFR per unit area within 
the active star forming disk.  The derived SFR surface densities are listed
in Table 1.  The observed \halpha\ surface densities (uncorrected for
extinction) can be derived from Table 1 by the simple relation: 
$ \log \Sigma_{H\alpha} = \log \Sigma_{SFR} + 34.65$, where 
$\Sigma_{H\alpha}$ is expressed in units of ergs~sec$^{-1}$~pc$^{-2}$.
This conversion may be useful for readers who may wish to apply a
different SFR calibration to the \halpha\ data compiled here.

%Table 1 lists for each of the 63 galaxies the derived SFR, atomic, 
%molecular, and total (atomic $+$ molecular) gas surface densities, and the 
%diameter used in the surface density calculation, as well as the relevant 
%references.  Somewhat larger samples of galaxies with HI or H$_2$ 
%measurements alone are available and are used in \S3, but the data are not 
%tabulated here, because the relevant information can be obtained elsewhere 
%(K89; Young \etal 1996).

\subsection{Infrared-Selected Starburst Galaxies}

The mean gas densities of the normal spiral disks in our sample
lie within a relatively narrow range,  
from 2 to 50 \msun~pc$^{-2}$, and this seriously 
limits the dynamic range over which the behavior of the Schmidt law
can be evaluated.  The density range can be extended to  
$\sim$100 \msun~pc$^{-2}$
by analyzing spatially-resolved measurements of individual disks 
(Paper II), but above these densities \halpha\ measurements become 
unreliable for determining the SFR.  For a typical gas-to-dust ratio 
found in nearby galaxies, the visual extinction reaches 1 mag for   
column densities $N_H \sim 2 - 4 \times 10^{21}$ cm$^{-2}$, or $\Sigma_H \sim  
15 - 30$ \msun~pc$^{-2}$ (e.g., Bohlin, Savage, \& Drake 1978; Caplan 
\& Deharveng 1986).  Hence one expects 
the extinction at \halpha\ to become problematic for regions with mean gas 
surface densities above 
50 -- 100 \msun~pc$^{-2}$.  If we wish to study the nature of the 
star formation law in these dense regions, a star formation diagnostic 
other than \halpha\ must be used.  

Large-scale star formation at much higher densities is commonly found in the
centers of normal galaxies, and particularly in luminous infrared
starburst galaxies.  In order to analyze the star
formation law in this regime, we searched the literature for 
high-resolution CO and infrared measurements of starburst galaxies. 
Since the starbursts are often concentrated in compact circumnuclear disks
(e.g., Scoville \etal 1994; Sanders \& Mirabel 1996; Smith \& Harvey 1996), 
high-resolution data are required in order to accurately determine the
linear sizes of the starburst regions and the corresponding surface
densities.  Our sample comprises 36 galaxies with 
high-resolution CO data, most based on aperture synthesis mapping, and 
for which infrared measurements of the same region are available.  The 
sample ranges from low-level nuclear starbursts in normal and peculiar 
galaxies such as NGC~253, IC~342, Maffei~2, and M82 
($L_{FIR} \sim 10^8 - 10^{10}$ \lsun) to 
ultraluminous starburst galaxies with $L_{FIR} > 10^{12}$ \lsun\ 
(e.g., Arp 220).  Care was taken to select objects in which the dust
heating is dominated by a starburst, as determined from optical spectra
spectra (e.g., Armus, Heckman, \& Miley 1989; Veilleux \etal 1995) 
and/or mid-infrared spectroscopy
(e.g., Lutz \etal 1996).  Objects with evidence for a strong AGN component
were excluded (e.g., NGC~1068, NGC~7469, Mrk 231, Mrk 273).  

Total molecular gas masses in the starburst disks 
were derived from the CO flux and distance, using the same 
CO/\h2\ conversion factor as for the normal galaxies.  The validity of 
a constant conversion factor is highly questionable (e.g., Wild 
\etal 1992; Downes, Solomon, \& Radford 1993; Aalto \etal 1994; 
Solomon \etal 1997), and we have adopted a uniform conversion factor
strictly for the sake of simplicity.  
The impact of adopting a different conversion
factor will be discussed later.  
The mean molecular surface densities were then derived, averaged 
within the radius of the central molecular disk as determined from the
CO maps.  

High-resolution HI observations are only available for a few of these 
galaxies, and in those cases the atomic fraction in the circumnuclear  
region is small, of order a few percent or less (e.g., 
Garcia-Barreto 1991; Downes \etal 1996; Sanders \& Mirabel 1996).
This is not surprising given the very high column densities found in
these regions.  Consequently we have 
ignored the HI component and approximate the molecular mass as 
the total gas mass in the starburst region.  Table 2 lists the galaxies
in the sample, the radii of the disks, and their mean molecular 
surface densities.

The SFRs for the starbursts were derived from measurements of their
FIR luminosities.  These were taken from a variety of sources, as listed 
in Table 2.  For about half of the sample, high-resolution maps at 
mid-infrared wavelengths are available, and when combined with IRAS 
fluxes for the galaxies as a whole they provide an accurate
estimate of the FIR luminosity in the central starbursts themselves
(Telesco, Dressel, \& Wolstencroft 1993; Smith \& Harvey 1996).  For 
the other galaxies the FIR luminosity of the starburst was
derived from a combination of IRAS photometry and groundbased aperture
photometry at 10--20 \microns, or from the IRAS fluxes alone, in cases where 
most of the total FIR emission
appears to originate in the central starburst.  SFRs for three of the
galaxies were derived from a combination of \brgamma\ and 
infrared photometry, as noted in Table 2.

In normal disk galaxies the relationship between the FIR luminosity and the 
SFR is complex, because stars with a variety of ages can contribute to the dust 
heating, and only a fraction of the bolometric luminosity of the young 
stellar population is absorbed by dust (e.g., Lonsdale \& Helou 1987; 
Walterbos \& Greenawalt 1996).  However in the starbursts studied here,
the physical coupling between the SFR and the IR luminosity is much more direct.
Young stars dominate the radiation field that heats the dust, and the 
dust optical depths are so large that 
almost all of the bolometric luminosity of the starburst is 
reradiated in the infrared.  This makes it possible to derive a reasonable
quantitative measure of the SFR from the FIR luminosity.  

Our calibration of the SFR/$L_{FIR}$ conversion is based on the starburst
synthesis models of Leitherer \& Heckman (1995).  Their models trace 
the temporal evolution of the bolometric luminosity for a fixed SFR,
metal abundance, and IMF.  We computed the SFR calibration using their
``continuous star formation" models, in which the SFR is presumed to
remain constant over the lifetime of the burst.  The models show that the
L$_{bol}$/SFR ratio evolves relatively slowly between ages of 10 and 100 Myr,
the relevant range for most of these starbursts  
(e.g., Bernl\"ohr 1993; Engelbracht 1997).  Alternatively one can derive
the conversion using a ``instantaneous burst" approximation, where it is 
assumed that star formation has ceased, but the calibration is sensitive
to the presumed burst age and the (questionable) assumption of an
instantaneous burst.  Adopting the mean luminosity for 10--100 Myr continuous
bursts, solar abundances, the Salpeter IMF described earlier,
and assuming that the dust reradiates all of the bolometric luminosity yields:
\begin{equation}
{ {SFR} \over {1~{\rm M_{\odot}~yr^{-1}}} } = { {L_{FIR}} \over {2.2 \times 10^{43}~ 
{\rm ergs~s^{-1}} } } =  
{ {L_{FIR}} \over {5.8 \times 
10^9~L_{\odot}} }
\end{equation}

\noindent
This lies within the range of previously published calibrations 
($1 - 3 \times 10^{-10}$~\msun~yr$^{-1}$~\lsun$^{-1})$.   
Equation (3) yields SFRs that are 14\%\ lower than the recent calibration of 
Lehnert \& Heckman (1996), and 22\%\ lower than Meurer \etal (1997).
The SFR surface density was then calculated within the radius of the 
starburst region as determined from the CO maps, or from the infrared
maps if high-resolution CO data were not available.  The sizes of the
regions defined in CO and the infrared show excellent correspondence
in cases where comparable resolution data are available
(Telesco \etal 1993; Smith \& Harvey 1996).  Table 2 lists the 
radii, gas densities, and SFR surface densities derived in this way. 

In \S4 we analyze the composite properties of the normal disk and 
starburst samples, so it is important to confirm that the FIR and 
\halpha-based SFRs are on a consistent zeropoint.  Matching aperture \brgamma\ 
photometry for 18 of the galaxies in our sample is available from the 
compilations of Puxley, Hawarden, \& Mountain (1990), Telesco \etal (1993)  
and Smith \& Harvey (1996), and these allow us to compare the emission-line
and FIR SFR scales 
on a self consistent basis.  The FIR-based SFRs were derived using 
equation (3), while the \brgamma-based SFRs were derived using equation 
(2) and a \brgamma/\halpha\ ratio of 0.0103, corresponding to Case B
recombination at $T_e = 7500$ K and $N_e = 10^3$ cm$^{-6}$ (Osterbrock 1989).  
No extinction corrections were applied to the \brgamma\ data.   
%For completeness Figure 1 also 
%shows with open triangles \brgamma\ and FIR fluxes for 8 other galaxies in our
%sample, taken from Goldader \etal (1997).  
%These data are not appropriate for the comparison here because the
%apertures used for the \brgamma\ 
%measurements are much smaller than for the FIR (IRAS fluxes were used).

Figure 1 shows a comparison of the FIR and \brgamma-derived SFRs.
The solid line shows the correlation expected
if the two sets of SFRs were equivalent.  The data in Figure 1 closely
follow this correlation, but the FIR-derived SFRs 
are systematically higher by an average of 0.29$\pm$0.06 dex,  
as shown by the dashed line.  This displacement could indicate a 
general inconsistency between the zeropoints of the \halpha\ and
FIR calibrations of the SFR, which might arise, for example, from errors
in the FIR luminosities (many of them extrapolated from the mid-IR), or
in the synthesis model that is used to convert the FIR luminosities to SFRs.
However there is physical justification for expecting that the \brgamma\ 
fluxes would systematically underestimate the SFRs in many of these objects.  
The extinction in most regions is so large that one expects part of
the ionizing radiation from the starburst to be absorbed by grains,
and in some objects extinction of \brgamma\ itself is probably significant
(e.g., Lutz \etal 1996; Goldader \etal 1997).
The \brgamma-derived SFR will also tend to be systematically lower
than the FIR-derived value if the starbursts are observed after the
peak of the burst, because the dust heating is dominated by longer lived
stars than the emission lines.  We provisionally adopt the 
SFRs from equation (3) in the following analysis, on the tentative 
assumption that the FIR-based SFRs are more reliable in these objects.
However we will also explore the consequences of adopting the lower 
\brgamma-based scale, and include this uncertainty in the analysis of 
the global Schmidt law.

\subsection{Uncertainties}

Individual uncertainties are not listed for the surface densities listed 
in Tables 1 and 2, because the predominant errors are systematic in nature and 
difficult to quantify.  However it is important to be aware of nature of these 
uncertainties and their possible influence on the observed star formation 
law.

For the normal spiral disks, with SFRs derived from \halpha\ 
luminosities (Table 1), the dominant systematic errors are extinction 
variations, which introduce a scatter in the SFR densities, and  
uncertainty in the extrapolated IMF, which could introduce an overall 
shift in the SFRs (Kennicutt 1983).  The dominant errors in the  
gas densities are variation in the CO/H$_2$ conversion factor, combined
with the limited sampling of the CO measurements in some galaxies  
(Sage 1993; Young \etal 1995).  A realistic estimate for the observational 
scatter in the SFRs is $\pm$30--50\%, or $\pm$0.15--0.3 dex 
(Kennicutt 1983), and the uncertainties in the gas densities are probably
comparable.  We adopt an average uncertainty of $\pm$0.2 dex in the
following analysis.

The systematic uncertainties in the SFRs and gas densities derived for 
the starburst galaxies (Table 2) are larger.  In many cases the FIR 
luminosities have been derived from a combination of high-resolution
mid-infrared measurements and IRAS FIR fluxes, and there can be substantial
uncertainty in the extrapolation to a total FIR flux.  In other cases
only integrated IRAS fluxes for the galaxies are available, and the 
presence of significant FIR emission from the region outside of the
central starburst will cause the starburst SFR to be systematically
overestimated.  The SFR will also be overestimated 
if the dust is heated partly by other sources, such as an active nucleus.
Another significant source of uncertainty in the SFRs inferred for 
individual starbursts is the use of a fixed continuous burst model,
though the effect on the overall SFR scale should be lower.
The gas densities in the starburst regions are also subject to 
systematic error as well, mainly through uncertainties in the CO/H$_2$ 
conversion factor (e.g., Downes \etal 1993; 
Solomon \etal 1997).  Other smaller sources of uncertainty include the
neglect of atomic gas and errors in the radii of the starbursts.  The 
latter errors affect the inferred SFR and gas densities equally, and 
have less of an effect on the form of the Schmidt law.

The largest of these systematic uncertainties, the L$_{FIR} -$ SFR conversion
and the CO/\h2\ conversion, could introduce errors in the SFR or gas
density scales at the factor of 2 $-$ 3 level (0.3 $-$ 0.5 dex).  In our
analysis we adopt uncertainties 
${^{+0.3}_{-0.5}}$ dex in both parameters, with the asymmetry reflecting 
the greater likelihood that the systematic errors tend to lead to overestimates
of the SFRs and gas densities.  Despite these uncertainties,
the data provide very strong constraints on the form of the 
star formation law, because of the very large range of absolute densities and 
SFRs represented in the sample, 2 $-$ 6 orders of magnitude depending on the 
subsample of interest.  
% However these uncertainties 
% should be borne in mind when following up on the analysis presented here.

\section{RESULTS}

\subsection{The Schmidt Law in Normal Disks}

Figure 2 shows the relationship between the disk-averaged SFR and
total gas density (atomic and molecular hydrogen)  
for the 61 normal spirals in our sample.  A clear correlation is apparent 
in the expected sense of increasing SFR with increasing gas densities, 
with a mean slope that is considerably steeper than a linear relation
(indicated by the dotted and dashed lines).
However the scatter in the relation is large, up to a 
factor of 30 in SFR at a fixed gas density, and comparable to the total 
range in observed gas density.  Consequently the slope of the Schmidt 
law is poorly constrained.   A conventional least squares fit 
which minimizes (logarithmic) residuals in the SFR density yields 
$N = 1.29 \pm 0.18$.
This slope lies in the middle of the range $N = 0.9 - 1.7$ derived in
previous studies with smaller samples 
(Buat \etal 1989; K89; Buat 1992; Deharveng \etal 1994).
A bivariate least squares regression, which takes into account the
uncertainties in the gas densities as well, yields a much steeper fit 
$N = 2.47 \pm 0.39$.  Both fits are shown with solid lines
in Figure 2.  The large difference between these solutions is a direct
reflection of the large dispersion in the disk-averaged SFR vs gas 
density relation, and the result underscores the conclusion that any 
Schmidt law in these galaxies should be regarded 
as a {\it very} approximate parametrization at best. 

What is the physical origin of the large dispersion in Figure 2?  As 
discussed earlier, variations in extinction and the CO/\h2\ conversion
introduce a scatter at about the $\pm$0.2 dex level in the
SFR and gas densities, as signified by the error
bars in Figure 2.  This can account for roughly half of the observed
scatter in the star formation law.  The remaining scatter
must be real, reflecting a real variation in the mean Schmidt law.
Such a variation is not entirely surprising, when one recalls that the 
local SFRs and gas densities span orders of magnitude within
typical disks, and averaging over the entire disk will not necessarily
preserve the form of a nonlinear local Schmidt law.
The problem is illustrated in Figure 3, which shows the radial SFR vs 
gas density profiles for 21 of the galaxies in our sample (Paper II).
Each profile was produced by measuring the 
azimuthally averaged gas density and SFR density as a function
of galactocentric radius, then plotting the resulting SFR vs gas
density relation on a common scale.
At high densities the SFRs are well represented by a shallow
Schmidt law ($N \sim 1.4$), but the slope of the star formation law
steepens abruptly below the threshold density.  The disk-averaged 
SFRs plotted in Figure 2 represent gross averages over these highly nonlinear
relations, and the resulting global Schmidt law exhibits a slope that is 
intermediate between the $N \sim 1.4$ power-law dependence at
high density and the steeper law in the threshold regime.  The dispersion
in Figure 2 is introduced because the star formation in some galaxies
is highly concentrated to the high-density part of the local Schmidt law,
while in other systems much of the star formation takes place near
the threshold density (see K89).  This 
underscores the caveat that disk-averaged Schmidt law analyzed here 
contains little physical information about the underlying star formation
law.  However it does provide a convenient means of parametrizing the
gross star formation properties of disks in simple one-zone evolution models.
We defer further discussion of the spatially-resolved star formation law
for Paper II.

The data in Figure 2 also provide useful information on the
average global efficiency of star formation in local disks, the
coefficient $A$ in equation (1).  The dashed and dotted lines in 
Figure 2 correspond to constant SFRs per unit gas mass, in units
of 1\%, 10\%, and 100\%\ per $10^8$ yr.  The choice of $10^8$ yr
as a fiducial timescale is arbitrary, but it does correspond roughly to 
a typical orbital time in the disks.  The lines are offset by
a factor of 1.37 to include helium and heavy elements in the 
total gas mass.  
The median efficiency for the disks in Figure 2 is 4.8\%, 
i.e., a typical present-day spiral galaxy converts 4.8\%\ of
the gas (within the optical radius) to stars over this period.  
%It is interesting that this global efficiency is comparable to the
%star formation efficiencies measured for typical individual molecular
%clouds in the Galaxy and elsewhere (e.g., Evans 199x).  It is  
%tempting to speculate that the global SFR in normal SFRs, at least in
%in gas-rich systems where large-scale threshold processes are unimportant, 
%may be set primarily by local physical processes on the scale of
%individual molecular clouds.  
The efficiencies 
can be expressed alternatively as gas consumption timescales, with
the lines in Figure 2 corresponding to timescales $\tau_{gas}$ of 10, 1, 
and 0.1 Gyr
(bottom to top).  The median gas consumption time for the disks in
this sample is 2.1 Gyr, again referring to the star forming disks alone,
and not including corrections for recycling of interstellar gas.
Recycling typically extends the actual consumption timescale 
by factors of 2--3 above the simple calculation (Kennicutt \etal
1994).

Most of the galaxies in Figure 2 possess disk-averaged star formation
efficiencies in the range 2 $-$ 10\%\ per $10^8$ yr, corresponding to
gas consumption times of 1 $-$ 5 Gyr.  However several galaxies are more
extreme, and the full range of efficiencies is 0.8 $-$ 60\%\ per $10^8$ yr 
($\tau_{gas}$ = 0.2 $-$ 12 Gyr).  The shortest timescales correspond
to optically-selected starburst galaxies such as NGC~1569 and NGC~3310, 
while the low extremes are represented by early-type spirals such as
M31, NGC~2841, and NGC~4698, where the current SFRs are so low that
the future consumption times, even for their modest gas supplies,
are comparable to the Hubble time. 

Until now our attention has focussed solely on the relationship
between the disk-averaged SFR and the total gas density, but we can
also examine how the SFRs correlate with the average atomic and molecular gas
densities, as shown in Figure 4.  These comparisons include
galaxies mapped in HI or CO (but not both), so the samples are considerably
larger than shown in Figure 2.  

The left panel of Figure 4 shows the SFR vs HI density relation for 88 galaxies
with \halpha\ and HI data in common.  The correlation is very reminiscent
of the SFR vs total density relation shown in Figure 2, and in fact
the correlation coefficients are nearly identical, 0.66 for the SFR $-$ HI
relation vs 0.68 for the SFR $-$ HI$+$\h2\ relation.  
This is not entirely surprising, as HI accounts for approximately
half of the total gas density on average.
These results are consistent with previous analyses based on smaller
samples by K89, Buat (1992), Deharveng \etal (1994), Boselli (1994),
and Boselli \etal (1995).  The physical
interpretation of the SFR vs HI Schmidt law is not obvious, however.  
It may trace the 
physical influence of the atomic gas density on the SFR, but it could
be that the SFR regulates the density of HI, through the photodissociation
of molecular gas by hot stars (Shaya \& Federman 1987; 
Tilanus \& Allen 1989).

The correlation between the \halpha-based SFRs and H$_2$ density is 
much weaker, as shown in the right panel of Figure 4.  This has
been reported previously, and appears to hold independently of whether
SFRs based on \halpha, UV continuum fluxes, or FIR fluxes are analyzed
(Buat 1992; Boselli 1994).  Such a poor correlation between the SFR and
molecular gas densities is unexpected, and it has led some to suggest 
that variations in the CO/H$_2$ conversion factor are responsible 
for the scatter (K89; Boselli 1994; Boselli
\etal 1995).  Our data provide indirect support for this interpretation.
Several lines of evidence suggest that the Galactic CO/\h2\ conversion
factor is valid in regions with near-solar metallicity, but that it tends to 
systematically
underestimate the H$_2$ mass in metal-poor regions, such as are found
in the outer disks of spirals or in low-luminosity galaxies
(e.g., Maloney \& Black 1988; Kenney \& Young 1988; Rubio \etal 1993; 
Wilson 1995).  To test whether 
this effect might be contributing to the scatter in Figure 4, we
subdivided our sample by blue luminosity, with solid points denoting
galaxies with L$_B > 10^{10}$ \lsun\ ($M_B < -19.5$ for  
H$_0$ = 75 km~s$^{-1}$~Mpc$^{-1}$) and open circles representing fainter
galaxies.  The mean metal abundance in disks is well correlated with
luminosity, so this provides an approximate separation of the galaxies 
by abundance, around a value of $\sim 1~Z_\odot$ (Zaritsky,
Kennicutt, \& Huchra 1994).  Figure 4 shows that the luminous, metal-rich 
spirals do show a much
better defined SFR vs H$_2$ density correlation, comparable in 
slope and scatter to the correlations with total and HI density.
By contrast, the low-luminosity galaxies show essentially no correlation
between the SFR and CO-inferred H$_2$ densities, with many CO-weak
galaxies showing unusually {\it high} SFRs.  Although this is hardly
a conclusive result, it offers circumstantial evidence 
that variations in the CO/H$_2$ conversion factor are responsible for
most of the scatter in the SFR vs molecular gas density relation.
Our conclusions are consistent with those of 
Boselli (1994) and Boselli \etal (1995), and the reader is referred
to those papers for more detailed discussions of this problem.

\subsection{The Schmidt Law in Circumnuclear Starbursts}

We can perform a parallel analysis for the infrared-selected starbursts,
and the results are summarized in Figure 5.  The comparison 
is directly analagous to that shown for the normal disks in 
Figure 2, except that the SFRs are derived from FIR luminosities, and 
the SFRs are correlated with the H$_2$ gas density alone (the
disks are expected to be overwhelmingly molecular, as discussed earlier).
The SFRs and densities are averaged within the radii of the central
molecular disks and starbursts, which have typical dimensions of 
order 1 kpc.  The error bars indicate the typical uncertainties, as
discussed in \S2.3.

The starburst galaxies also show a well-defined Schmidt law, in this case
with a best fitting least squares slope $N = 1.40 \pm 0.13$ (bivariate
regression) or $N = 1.28 \pm 0.08$ (errors in SFRs only).
The Schmidt law is better defined than for the normal
disks, but partly because there is a much larger dynamic range in SFR and gas 
densities in the starburst sample; the dispersion in absolute SFR per
per unit area at fixed gas density is only slightly lower in the starburst
sample.  Star formation threshold effects are probably unimportant in
the starburst disks, and this might also account for the somewhat
tighter Schmidt law among these objects.

Although the starburst disks exhibit a SFR vs gas density relation that
is qualitatively similar in form to that seen in the normal spiral disks, the
physical regime we are probing is radically different.  The average gas 
surface densities 
here range from $10^2$ to $10^5$ \msun~pc$^{-2}$, compared to a typical
range of order 1 $-$ 100 \msun~pc$^{-2}$ in normal disks (Figures 2, 3).  
The mean densities of the starburst disks are comparable instead to those of
individual molecular cloud complexes in normal galaxies.
For example, the largest HII/GMC complexes in M31, M33, and M51 have
molecular masses and sizes corresponding to mean surface densities of
40 $-$ 500 \msun~pc$^{-2}$ (Wilson \& Rudolph 1993; Wilson \& Scoville 1992;
Nakai \& Kuno 1995).  This is comparable to the {\it low end} of  
the density range for the starbursts in Figure 5.  The mean densities
of some of the starbursts approach those of Galactic molecular cloud cores,
but extending over kiloparsec diameter regions.  The star formation densities
are just as extraordinary.  For example, the central 10 pc core of 
the 30 Doradus giant HII region contains $\sim 10^4$ \msun\ in young stars,
which corresponds to $\Sigma_{SFR} \sim 100$ \msun~yr$^{-1}$~kpc$^{-2}$ if
the star formation timescale is as short as $10^6$ yr;  
the average SFR density averaged over the entire HII region is 
$\sim$1 $-$ 10 \msun~yr$^{-1}$~kpc$^{-2}$.  
Thus the regions we are studying
have projected SFRs per unit area that approach the maximum limit
observed in nearby optically-selected star clusters and associations
(Meurer \etal 1997), but extending over regions up to a kiloparsec in
radius.

Not surprisingly, the global star formation efficiencies in the starburst
sample are much higher than in the normal disk sample (e.g., Young \etal 
1986; Solomon \& Sage 1988; Sanders, Scoville, \& Soifer 1991).
In Figure 5 we show the same lines of constant star formation efficiency
and gas consumption times as in Figure 2 (1\%, 10\%, and 100\%\ per $10^8$
yr).  The median rate of gas consumption is 30\%\ per 10$^8$ yr,
6 times larger than for the normal disk samples, and the efficiencies
reach 100\%\ per $10^8$ yr for the most extreme objects.  It is 
interesting to note that the shortest gas consumption times are comparable 
to the dynamical timescales of the parent galaxies, implying that the most
luminous starbursts are forming stars near the limit set by the 
gas accumulation timescale (Lehnert \& Heckman 1996).

\section{THE COMPOSITE SCHMIDT LAW}

Taken together, the normal disk and starburst samples span a dynamic
range of approximately 
$10^5$ in gas surface density and over $10^6$ in SFR per unit 
area.  Figure 6 shows the composite relation, with the normal spirals
shown as solid circles and the starbursts as solid squares.  
% Two plots are shown; the left panel combines the data in Figures 2 and 5
% directly, while in the right panel the SFRs for the starbursts have
% been reduced by 0.30 dex to place them on the zeropoint of the \brgamma\
% data (\S2.2).  
Quite remarkably, the data are consistent with a common 
Schmidt law extending over the entire density range.

Figure 6 shows that the normal disk and starburst samples occupy
completely separate regimes in gas density and SFR per unit area,
not a surprising result given the very different selection criteria
for the two samples.  But before we interpret the composite relation it is 
important to establish whether there is a smooth physical continuity between
the normal disk and starburst regimes, and to confirm the consistency 
of the \halpha\ and FIR-derived SFR scales.  To this end we derived 
\halpha-based SFRs and gas densities for the central regions of 
25 of the normal spirals in Table 1 ($R < 25$\arcsec), using our 
\halpha\ images and published HI and CO maps (Paper II).  The resulting
SFR and gas densities are shown as open circles in Figure 6.
These regions span the physical parameter space between the normal
disks as a whole and the infrared-selected circumnuclear starburst regions.
Figure 6 shows that the gas densities and \halpha-derived SFRs of these
regions fall on the composite Schmidt law defined by the normal disk
and starburst samples, and fill the transition region between the two
physical regimes.  The same conclusion can be drawn by comparing the
SFRs of the infrared-selected starburst galaxies in Figure 5 with the
spatially-resolved SFRs of the normal disks shown in Figure 3; the
starbursts lie on the extrapolation of the high-density star formation
laws observed in the spiral disks.  
This result, combined with the \brgamma-FIR comparison
discussed earlier, gives us confidence that we are
measuring the form of the star formation law on a self-consistent basis
across the sample.

The solid line in Figure 6 shows a bivariate least-square fit to the composite
relation defined by the normal disks and the starbursts (but not including
the open circles).  In this 
case we applied equal weights to all of the data points, in order to
avoid having the fit driven by the normal spirals in the lower left
region of Figure 6.  This yields a best fitting index $N = 1.40 \pm 0.05$
(bivariate regression) or $N = 1.35 \pm 0.03$ (errors in SFRs only).
These are nearly identical to the Schmidt law fits for the starburst
sample alone, which further confirms the consistency of the large-scale
star formation laws in the two samples.

The formal uncertainties listed here assume random
errors of $\pm$0.3 dex in the SFRs and gas densities, but it 
underestimates the full uncertainty in the Schmidt
law, because we have not accounted for the possibility of a
systematic shift in the overall SFR or density scales for the starburst
sample as a whole.  The effect of such a shift is easily calculated. 
For example, reducing the SFRs for all of
the starbursts by a factor of two, to match the \brgamma\ calibration in
Figure 1, would lower the best fitting index $N$ from 1.40 to 1.28.  
Likewise, lowering the gas masses in the starbursts by a factor of two, 
to take into account the possibility that the CO/\h2\ conversion factor is 
systematically lower, would {\it increase} $N$ 
by approximately the same amount, from 1.4 to 1.5.  This range of values
provides a fairer estimate of the actual uncertainty in the composite
Schmidt law.  Folding together all of these uncertainties, 
we adopt as our final result:
\begin{equation}
 {\Sigma_{SFR}} = {(2.5 \pm 0.7) \times 10^{-4}}~{{ ({{\Sigma_{gas}} \over 
 {1~M_{\odot}~{\rm pc}^{-2}}})}^{1.4 \pm 0.15}}~
  {{\rm M_{\odot}~yr^{-1}~kpc^{-2}}}  .
\end{equation}

Figure 6 shows that equation (4) provides an excellent parametrization 
of the global SFR, over a density range
extending from the most gas-poor spiral disks to the cores of the most
luminous starburst galaxies.  This may account for why 
conventional galaxy evolution models, which usually are based on a 
Schmidt law parametrization of the SFR, often produce realistic 
predictions of the gross star formation properties of galaxies.

There are limitations to the Schmidt law in equation (4) that should
be borne in mind, however, 
when applying this recipe to galaxy evolution models
or numerical simulations.  Although the full range of SFRs and gas densities
are very well represented by a single power law with $N \simeq 1.4$,
the scatter in SFRs about the mean relation is substantial,
$\pm$0.3 dex rms, and individual galaxies deviate by as much as
a factor of 7.  Consequently equation (4) provides at most a statistical 
description of the global SFR, averaged over large samples of galaxies.
Another potential limitation for its application to simulations and  
models is the need to accurately specify the linear sizes of
the relevant star forming regions.  This is relatively straightforward
for normal disks, where the scaling radius is comparable to the photometric
radius of the galaxy or the edge of the active star forming disk.
It may be more difficult to model in starbursts, however, where the 
intense star formation is usually concentrated in a region that is a 
few percent of the radius of the parent galaxy.  Fortunately 
the slope of the Schmidt law is relatively shallow,
and a modest error in the scaling radius will displace the inferred SFR and
gas densities nearly along a line of slope $N = 1$, nearly parallel
to the Schmidt law itself.  This is illustrated 
in Figure 6, where a short diagonal line shows the effect of changing
the scaling radius by a factor of two (for a fixed gas mass and total SFR).

\section{DISCUSSION: INTERPRETATION AND OTHER RECIPES}

The Schmidt law in Figure 6 is so well defined that it is 
tempting to identify a simple, unique physical origin for the relation.
However we find that a Schmidt law is not the only simple parametrization
that can reproduce the range of SFRs observed in this sample, and this
serves as a caution against overinterpreting the physical nature of 
the empirical star formation law.  In this section we briefly discuss
the form of the Schmidt law expected from simple gravitational arguments,
and demonstrate that a simple kinematical model 
provides an equally useful recipe for modelling the large-scale SFR.

Numerous theoretical scenarios which produce a Schmidt law 
with $N$ = 1 $-$ 2 can be found in the literature 
(Larson 1992 and references therein).  Simple self-gravitational
models for disks can reproduce the large-scale star formation thresholds
observed in galaxies (Quirk 1972; K89), and the same basic model is consistent
with a Schmidt law at high densities with index $N \sim 1.5$ 
(Larson 1988, 1992).  For example in a simple self-gravitational picture
in which the large-scale SFR is presumed to scale with the growth rate of 
perturbations in the gas disk, the SFR will scale as
the gas density divided by the growth timescale: 
\begin{equation}
{\rho_{SFR}} \propto { {\rho_{gas}} \over {{(G \rho_{gas})}^{-0.5}}} \propto
 {{\rho{_{gas}}^{1.5}}}  .
\end{equation}

\noindent
where $\rho_{gas}$ and $\rho_{SFR}$ are the volume densities of 
gas and star formation.  The corresponding scaling of the projected surface
densities will depend on the scale height distribution of the gas, with
$N = 1.5$ expected for a constant mean scale height, a 
reasonable approximation for the galaxies and starbursts considered here.
Although this is hardly a robust derivation, it does show that
a global Schmidt law with $N \sim 1.5$ is physically plausible.

In a variant of this argument, Silk (1997) has suggested a
generic form of the star formation law, in which the SFR surface 
density scales with the ratio of the gas density to the local dynamical
timescale:
\begin{equation}
\Sigma_{SFR} \propto { {\Sigma_{gas}} \over {\tau_{dyn}}} \propto 
   {\Sigma_{gas}~\Omega_{gas} }
\end{equation}

\noindent
where $\tau_{dyn}$ refers in this case to the local orbital timescale
of the disk, and $\Omega$ is the angular rotation speed.  Models of this
general class have been studied previously by Wyse (1986) and Wyse \&
Silk (1989), though with different scalings of the gas density and 
separate treatment of the atomic and molecular gas.   Equation (6)
might be expected to hold if, for example, star formation triggering
by spiral arms or bars were important, in which case the SFR would scale 
with orbital frequency.  To test this idea,   
we compiled rotation velocities for the galaxies in Tables 1 and 2, and
used them to derive a characteristic value of $\tau_{dyn}$ for each disk.
The timescale $\tau_{dyn}$ was defined arbitrarily as 
$ {2 \pi R / V(R)} = 2 \pi / \Omega(R)$, the orbit time at the outer radius 
$R$ of the star forming region.  The mean orbit time in the star forming 
disk is smaller than $\tau_{dyn}$ defined in this way, 
by a factor of 1 $-$ 2, depending on the form of the 
rotation curve and the radial distribution of gas in the disk.  We chose to
define $\tau_{dyn}$ and $\Omega$ at the outer edge of the disk to avoid these 
complications.  Tables 1 and 2 list the adopted values, in units of $10^8$ yr.
Face-on galaxies or those with poorly determined (rotational) velocity
fields were excluded from the analysis.

Figure 7 shows the relationship between the observed SFR density
and $\Sigma_{gas} / \tau_{dyn}$ for our sample.  The solid line
is not a fit but simply a line of slope unity which bisects
the relation for normal disks.  This alternate prescription
for the star formation law provides a surprisingly good fit to
the data, both in terms of the slope and the relatively small scatter
about the mean relation.  When compared over the entire density range
the observed law is slightly shallower than predicted by equation (7)
(slope $\sim$0.9 instead of 1); on the other hand the fit to the normal
disk sample is as tight as a Schmidt law.  The zeropoint of 
the line corresponds to a SFR of 21\% of the gas mass per
orbit at the outer edge of the disk.  Since the average orbit
time within the star forming disk is about half that at the disk
edge, this implies a simple parametrization of the local star 
formation law:
\begin{equation}
\Sigma_{SFR} \simeq 0.017~\Sigma_{gas}~\Omega_{gas}   ,
\end{equation}

\noindent
in other words the SFR is $\sim$10\%\ of the available gas mass 
per orbit.  

From a strictly empirical point of view, the Schmidt law in equation (4)
and the kinematical law in equation (7) offer two equally valid 
parametrizations for the global SFRs in galaxies, and either can be
employed as a recipe in models and numerical simulations.  It is
unclear whether the kinematic model can fit the radial distribution
of star formation as well as a Schmidt law, and we plan to explore
this in Paper II.

The two parametrizations also offer two distinct interpretations of 
the observation that the star formation efficiency in
central starbursts is much higher than found in quiescent star forming
disks (e.g., Young \etal 1986; Solomon \& Sage 1988; Sanders \etal 1991).  
In the Schmidt law picture, the higher efficiencies in
starbursts are simply a consequence of their much higher gas densities.
For a given index $N$, the SFR per unit gas mass will scale as 
$\Sigma{_{gas}^{(N - 1)}}$, and hence for the law observed here 
roughly as $\Sigma{_{gas}^{0.4}}$.  The
central starbursts have characteristic gas densities that are 
100 $-$ 10000 times higher than the average for normal disks, hence we 
would expect the global star formation efficiencies to be 6 $-$ 40 times
higher, as observed. 
In the alternative picture in which the SFR is presumed to scale with 
$\Sigma_{gas} / \tau_{dyn}$, the high SFRs and star formation efficiencies
in starburst galaxies simply 
reflect the smaller physical scales and shorter dynamical timescales
in these compact central regions.  It is difficult to 
to differentiate between these alternatives with disk-averaged measurements
alone, and since the global star formation law is mainly
useful as an empirical parametrization, the distinction may not be important.
Deeper insight into the physical nature of the star formation law requires
spatially resolved data for individual disks, of the kind that will
be analyzed in Paper II.

\acknowledgments
Several individuals contributed to the large set of \halpha\ data 
analyzed in this paper, and it is a pleasure to thank them.  The KPNO
data used in this paper were obtained as part of other projects in 
collaboration with R. Braun, R. Walterbos, and P. Hodge.  C. Martin
worked on the reduction of the spatially resolved \halpha\ data
shown in Figure 3.  I am also grateful to 
J. Black, J. Ostriker, S. Sakai, P. Solomon, S. White, and especially J. Silk 
for comments and suggestions about early versions of this work.  I am
also grateful to the anonymous referee for several comments that
improved the paper.
Some of the data used in this paper were obtained on the 2.3m Bok
telescope at Steward Observatory.  
This research was supported by the National Science Foundation
through grant AST-9421145.

\newpage

{\bf Figure Captions}

FIG. 1.--- A comparison of integrated SFRs derived from \brgamma\
emission-line fluxes and far-infrared continuum luminosities, for 18
infrared-selected starburst galaxies.  The solid line shows the 
relation expected from eqs. (2) and (3).  The dashed line is the
best fitting mean relation.

FIG. 2.--- Relation between the disk-averaged SFR per unit area
and gas density for 61 normal disk galaxies.  The solid lines are  
least square fits to the Schmidt law, as described in the text.
The dashed and dotted
lines correspond to constant global star formation efficiencies
and gas consumption timescales, as indicated.

FIG. 3.--- Profiles of the azimuthally averaged SFR per unit area
as a function of gas density for 21 spirals with spatially resolved
\halpha\ data.  

FIG. 4.--- Correlation of the disk-averaged SFR per unit area
with the average surface densities of HI (left) and \h2\ (right).
The \h2\ densities were derived using a constant CO/\h2\ conversion
factor.  In the right panel, solid circles denote galaxies with
L$_B > 10^{10}$ \lsun, while open circles denote galaxies with 
L$_B < 10^{10}$ \lsun. 

FIG. 5.--- Relation between the disk-averaged SFR per unit area
and molecular gas density for 36 infrared-selected circumnuclear 
starbursts.  The solid line shows a bivariate least squares fit 
to the Schmidt law,
as described in the text.  The dashed and dotted lines correspond to
constant global star formation efficiences and gas consumption timescales,
as indicated.

FIG. 6.--- Composite star formation law for the normal disk (solid
circles) and starburst (squares) samples.  Open circles show the SFRs
and gas densities for the centers of the normal disk galaxies.
The line is a least squares fit with index $N = 1.40$.  The diagonal
short line shows the effect of changing the scaling radius by a factor of 
two.

FIG. 7.--- Relation between the SFR for the normal disk and starburst
samples and the ratio of the gas density to the disk orbital timescale,
as described in the text.  The symbols are the same as in Figure 6.
The line is a median fit to the normal disk sample, with the slope
fixed at unity as predicted by equation (7).

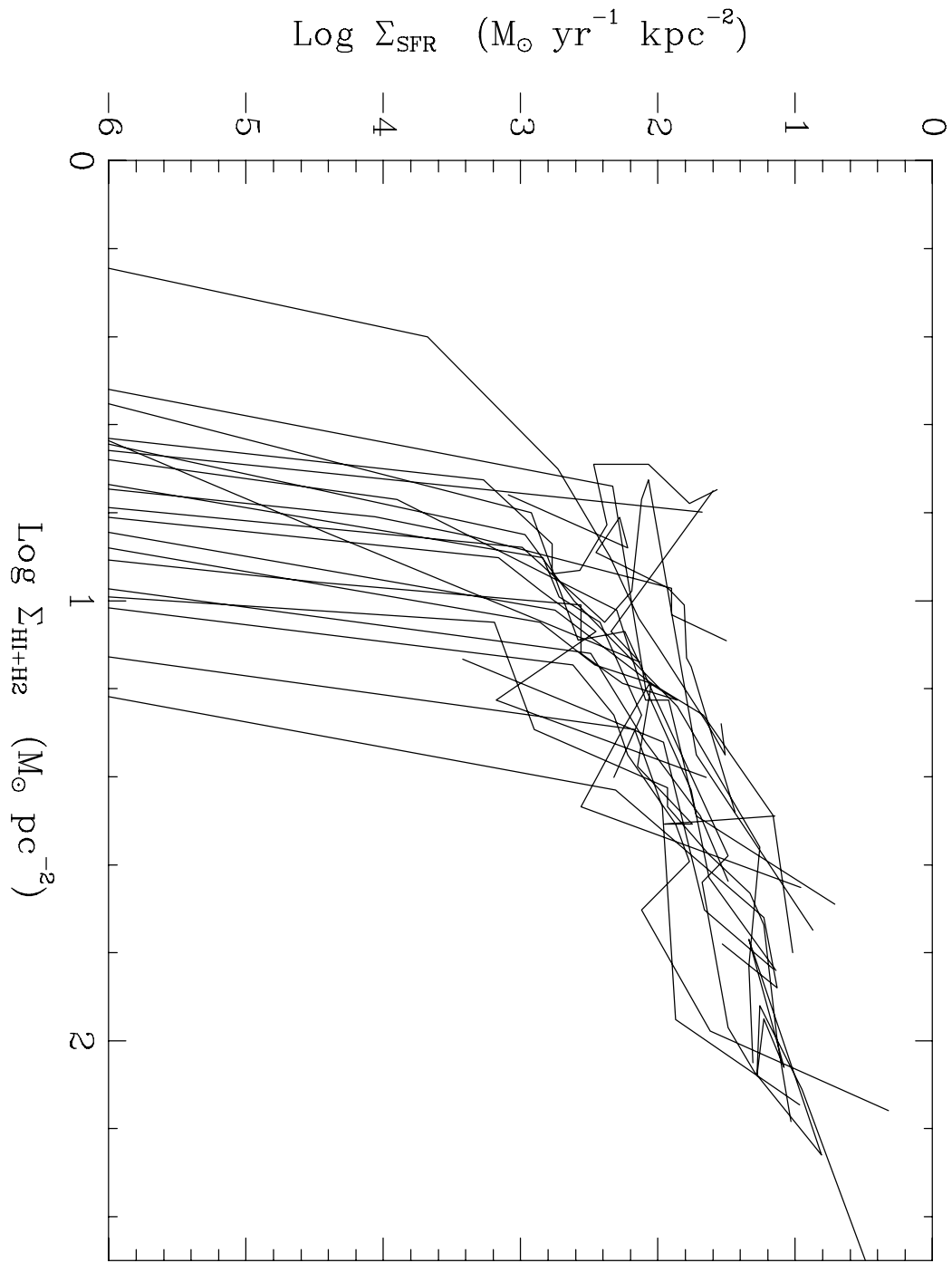
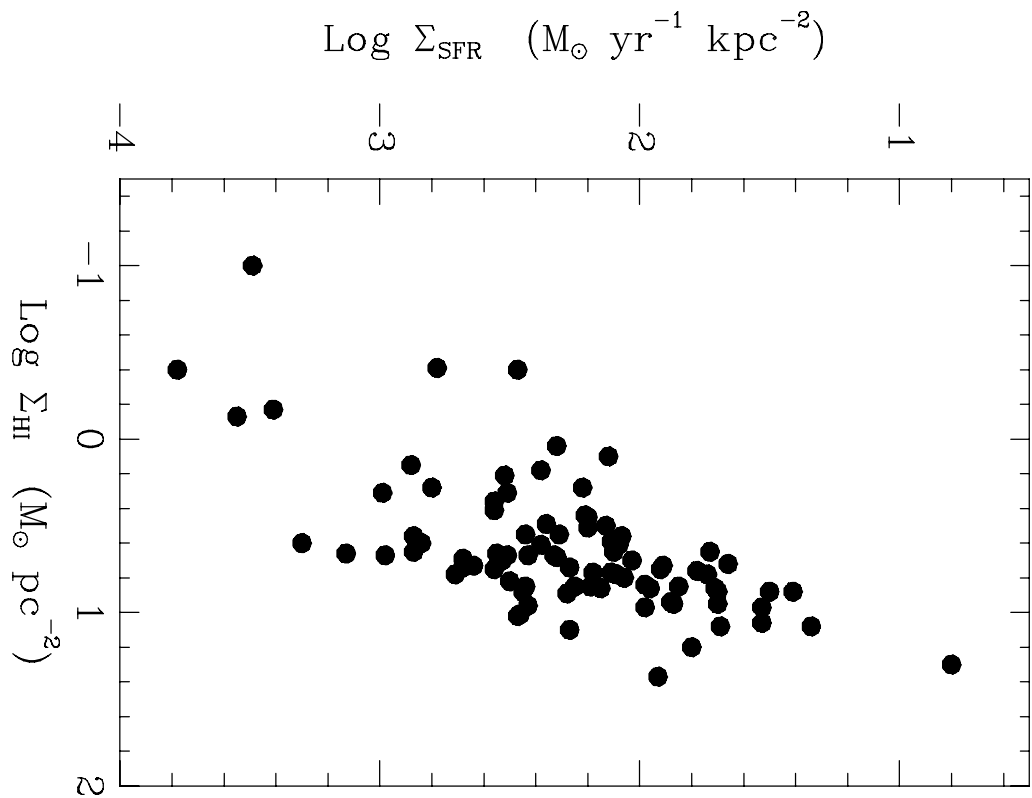
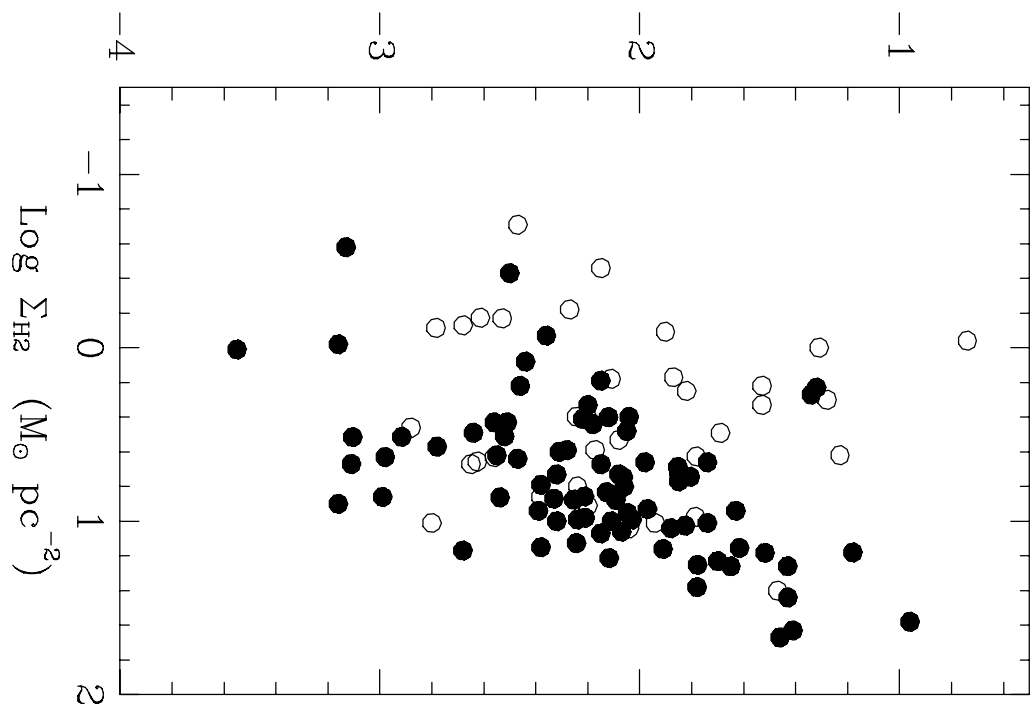
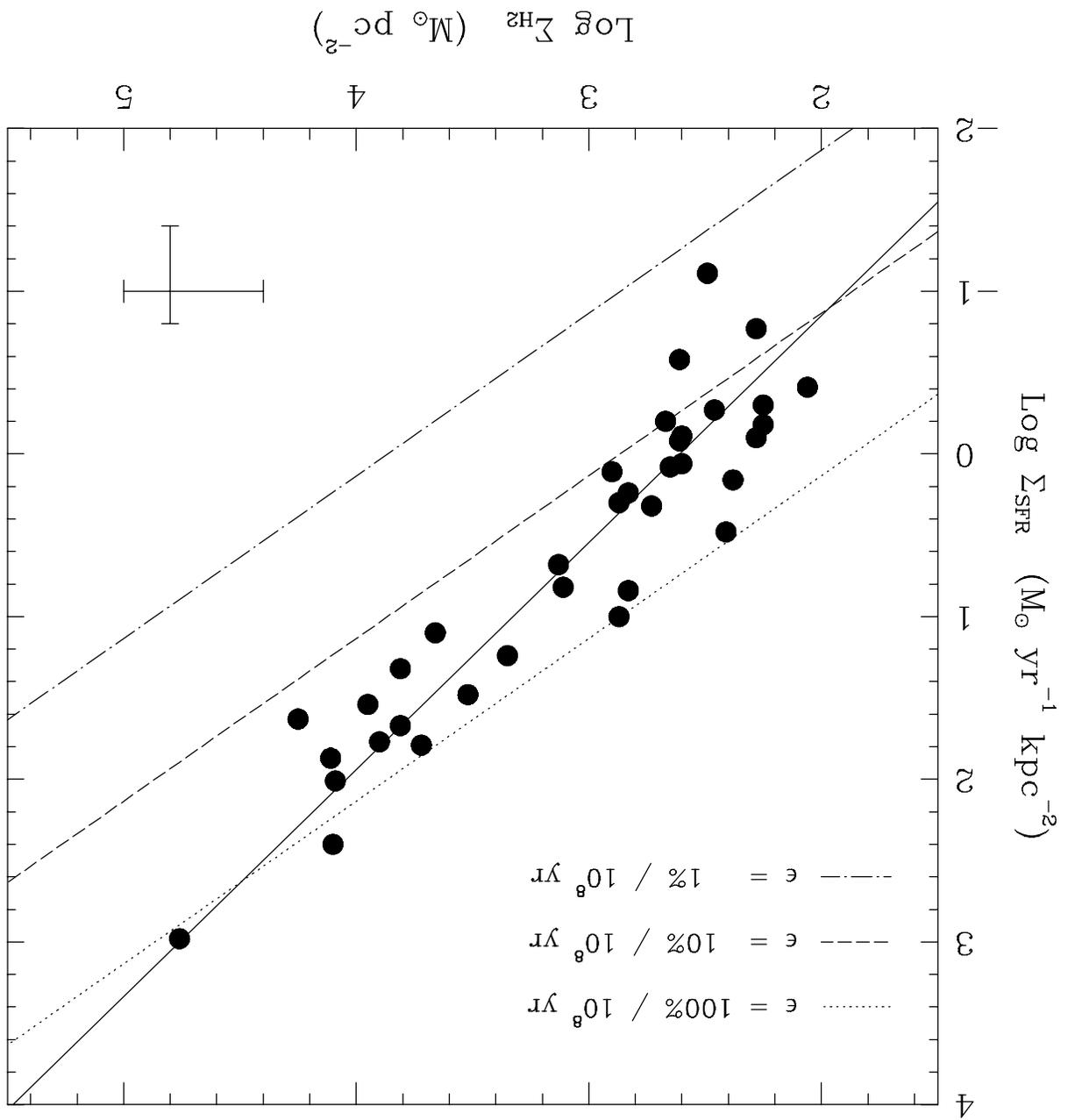
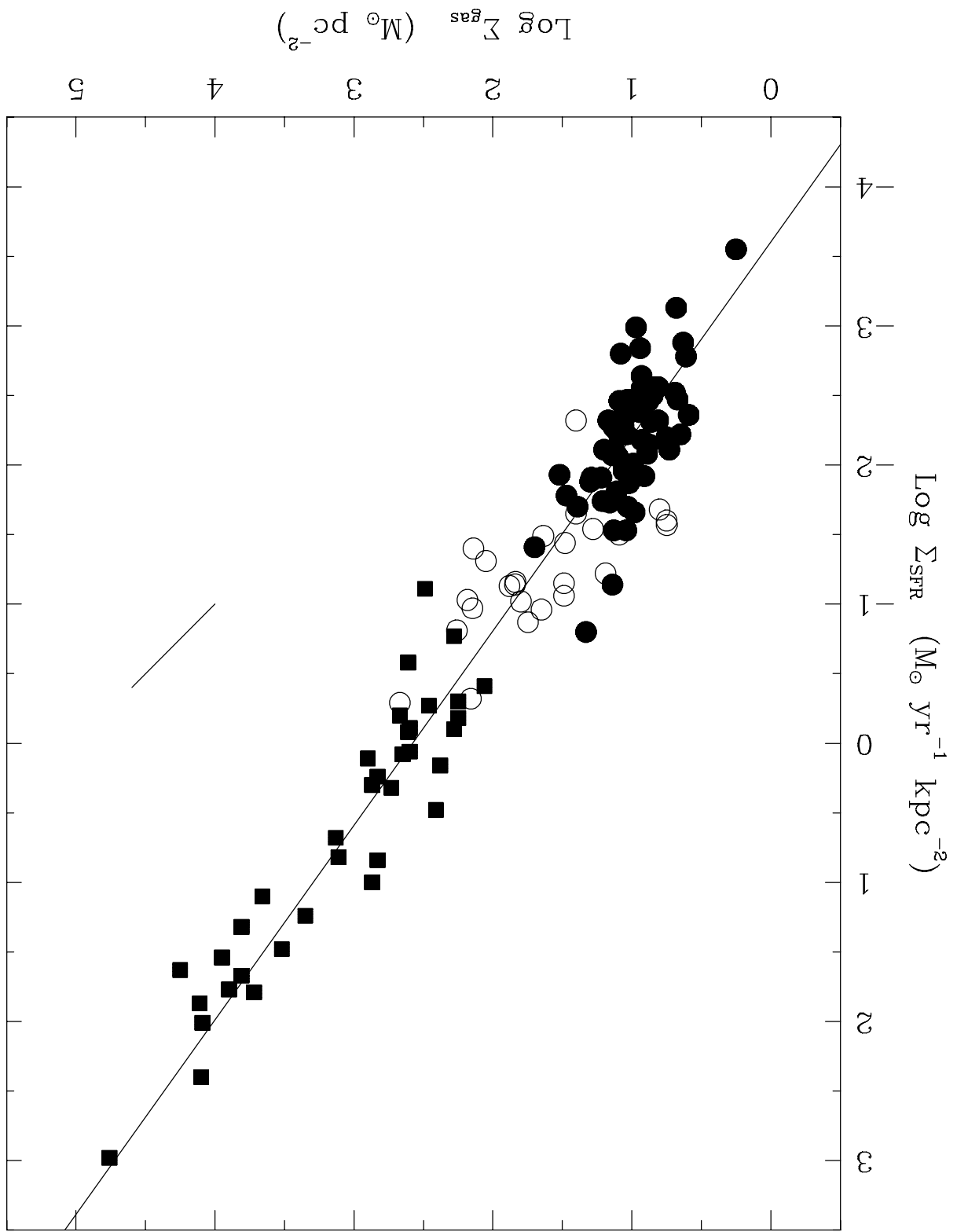
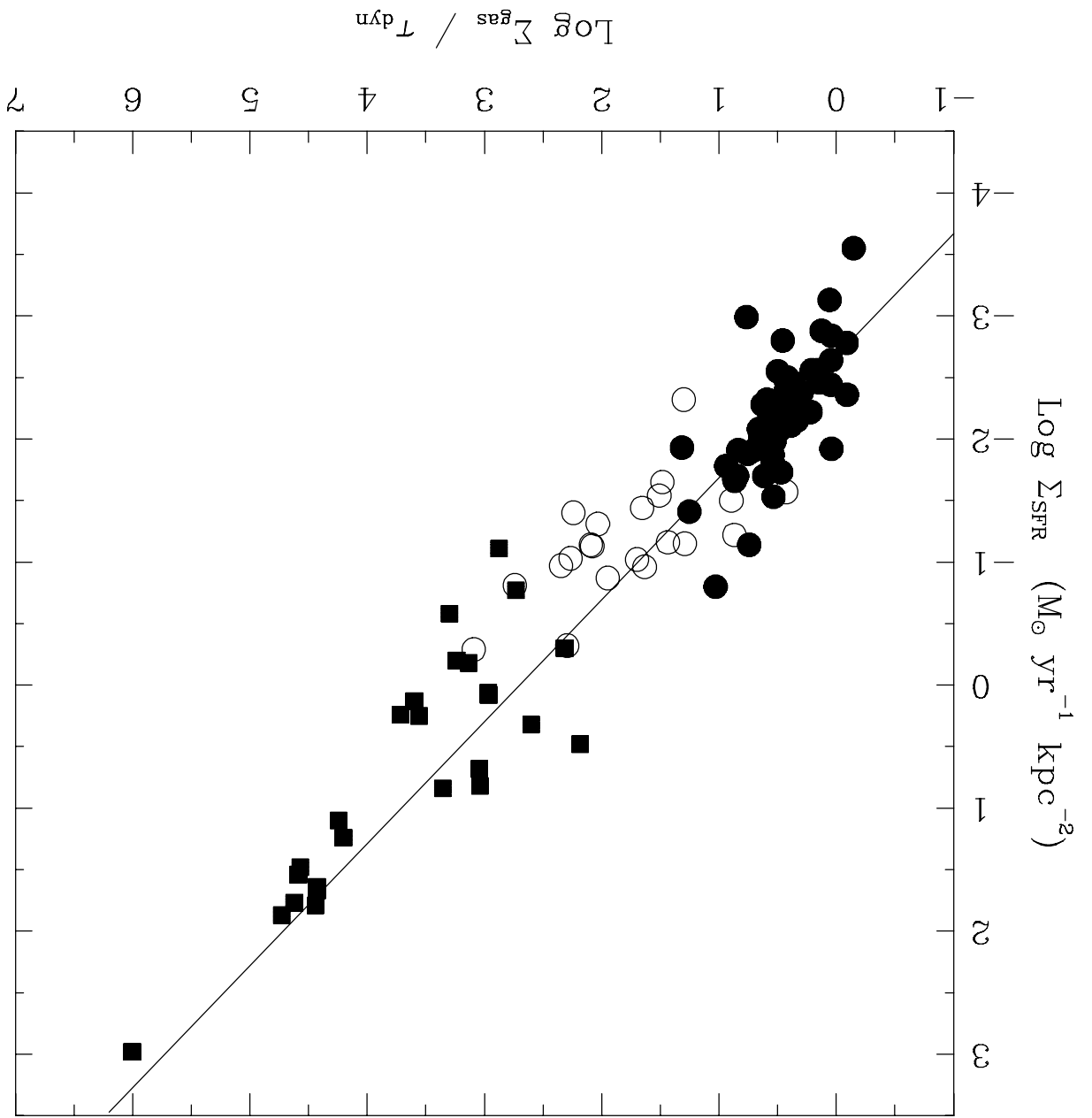
\end{document}